\documentclass[prl,twocolumn,superscriptaddress]{revtex4}
\usepackage{units,bm}
\usepackage{graphicx}
\usepackage{psfrag}

\newcommand{\ve}[1]{\Vec{#1}}
\newcommand{\Vec}[1]{\bm{\mathbf{#1}}}
\newcommand{\diff}[1]{\mathop{\mathrm{d}#1}\nolimits}

\begin{document}

\title{Lorentz invariance and quantum gravity: an additional
  fine-tuning problem?} 
\author{John Collins}
\author{Alejandro Perez}
\affiliation{Physics Department,
Pennsylvania State University,
University Park, PA 16802, USA
}
\author{Daniel Sudarsky}
\affiliation{Physics Department,
Pennsylvania State University,
University Park, PA 16802, USA
}
\affiliation{Instituto de Ciencias Nucleares,
Universidad Nacional Aut\'onoma de M\'exico,
A. Postal 70-543, M\'exico D.F. 04510, M\'exico
}
\author{Luis Urrutia}
\affiliation{Instituto de Ciencias Nucleares,
Universidad Nacional Aut\'onoma de M\'exico,
A. Postal 70-543, M\'exico D.F. 04510, M\'exico
}
\author{H\'ector Vucetich}
\affiliation{Instituto de F\'\i sica,
Universidad Nacional Aut\'onoma de M\'exico
A. Postal 70-543, M\'exico D.F. 04510, M\'exico
}
\affiliation{Observatorio Astron\'omico,
Universidad de La Plata,
La Plata, Argentina
}

\date{30 October 2004}


\begin{abstract}
  Trying to combine standard quantum field theories with gravity leads
  to a breakdown of the usual structure of space-time at around
  the Planck length, $\unit[1.6\times10^{-35}]{m}$, with
  possible violations of Lorentz invariance.
  Calculations 
  of preferred-frame effects in quantum gravity have further motivated
  high precision searches for
  Lorentz violation.
  Here, we explain that
  combining known elementary particle interactions with a Planck-scale
  preferred frame gives rise to Lorentz violation at
  the percent level, some 20 orders of magnitude higher than
  earlier estimates, unless the bare parameters of the theory are
  unnaturally strongly fine-tuned.
  Therefore an important task is not just the improvement
  of the precision of searches for violations
  of Lorentz invariance, but also the search for theoretical mechanisms for
  automatically preserving Lorentz invariance.
\end{abstract}



\maketitle


The need for a theory of quantum gravity and a modified structure of
space-time at (or before) the Planck scale is a consequence of the
known and successful theories of classical general relativity (for
gravity) and the standard model (for all other known interactions).
Thus one of the most important challenges in theoretical physics is
the construction of a quantum theory of gravitation.

Direct investigations of Planck-scale phenomena need short-wavelength
probes with elementary-particle energies of order the Planck energy
$E_P = (\hbar c^5 / G)^{1/2} = \unit[1.2\times10^{19}]{GeV}$, which is
much too high to be practicable.  But actual
tests --- e.g.,
\cite{Coleman.Glashow,signal.CK,tests}
---
of a hypothesized
granularity of space-time at the Planck scale are possible because
relativity (embodied mathematically as Lorentz invariance) gives a
unique form for the dispersion relation between the energy and
momentum of a particle,
\begin{equation}
  \label{eq:DR.normal}
  E = \sqrt{\ve{p}^2c^2 + m^2c^4}.
\end{equation}
Here $c$, the speed of light is a universal constant, while the
particle rest mass $m$ depends on the kind of particle.
We will henceforth use units in which $c=1$.

Calculations in
\cite{calcs,lqg1}
find
preferred-frame effects associated with space-time granularity
\cite{footnote}
\nocite{Kozameh:2003rm,ARMU}
in the two most popular contenders for a theory of quantum
gravity, which are string theory \cite{string.review2} and loop
quantum gravity \cite{lqg.review1,lqg.review2}.  In these
scenarios, the preferred frame and the consequent Lorentz
violation occur even though the fundamental classical equations of
both of the theories are locally Lorentz invariant.  
We thus have
a quantum inspired revival of the nineteenth century idea of the
electromagnetic ether, a background in which propagate light
waves, as well as all other elementary particles and fields.
Specific estimates of modified dispersion relations were made in
these papers from calculations of the propagation of quantum
mechanical waves in the granular space-time background.  At
accessible energies, only minute effects were predicted, of
relative order $E/E_P$ or $(E/E_P)^2$, when the probe has energy
$E$.  
For other ways in which Lorentz violation might arise,
see, for example, \cite{Burgess:2002tb,ABDG}.
The minuteness of the effects is in accord with everyday
scientific thinking, where we often find that the details of
physical phenomena on one distance scale do not directly manifest
themselves in physics on much larger scales. Therefore attention
has focused on 
searches for extremely
small violations of the dispersion relation.

\begin{figure}
\centering
\psfrag{k}{$k$}
\psfrag{p}{$p$}
\includegraphics[height=2cm]{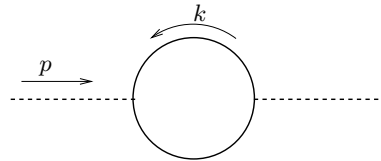}
\caption{Lowest order self-energy graph.
  Interactions of quantum
  fields require an unrestricted integral over the momenta of the
  virtual particles up to the highest momenta allowed in the theory. }
\label{fig:loop}
\end{figure}

However, as we will now explain, the
predicted violations of the dispersion relations are enormously
increased when we include known elementary particle interactions.
In quantum field theories like the standard model, the propagation
of an isolated particle has calculable contributions
from Feynman diagrams for particle
self-energies, such as Fig.\ \ref{fig:loop}.  The dispersion law
for a particle is obtained by solving
\begin{equation}
  \label{eq:inv.prop}
  E^2 -\ve{p}^2 - m^2 - \Pi(E,\ve{p}) = 0.
\end{equation}
Here $\Pi$
is the sum of all self-energy graphs, to which we have added any (small)
Lorentz-violating corrections calculated in free-field theory as in
\cite{calcs,lqg1}.

We now apply the following reasoning: Without a cutoff the graphs
have divergences from large momenta/short distances.  In the
Lagrangian defining the theory, the divergences correspond to
terms of dimension 4 (or less) that obey the symmetries of the
microscopic theory.  In the textbook situation with Lorentz
invariance, the divergences are removed by renormalization of the
parameters of the theory.  Now Planck-scale physics can cut off
the divergences, and also modify the formulae to be used when loop
momenta are around the Planck scale.  But the same power-counting
that determines the divergences also determines the natural size
of the contributions of Planck-scale virtual momenta: the dominant
contributions correspond to operators of dimension 4 (or
less) in the Lagrangian.  (This is the argument that leads to the
well-known concept of an effective low-energy theory.)  If the
microscopic theory violates Lorentz invariance at the Planck
scale, then generically we also get Lorentz violation at low
energies without any suppression by powers of $E/E_P$; for some
details, see the Appendix.  This Lorentz violation can be removed
by explicit Lorentz-violating counterterms of dimension 4 in the
Lagrangian which are fine-tuned to give the observed low-energy
Lorentz invariance.  But such fine-tuning is unacceptable
\cite{Weinberg,fine.tuning1,fine.tuning2} in a fundamental theory.

In the scenarios of
\cite{calcs,lqg1},
Lorentz
violating corrections to the dispersion relation, suppressed by
one or two powers of $E/E_P$, were found by considering the
propagation of \emph{free} particles in a granular space-time
background.
The above reasoning shows, instead, that there are effects that
are only suppressed by two powers of known standard-model
couplings, effects that change the value of $c$ in the dispersion
relation (\ref{eq:DR.normal}).  The known, widely different,
couplings of different elementary fields imply that different
fields have different values of $c$, with fractional differences
around
$0.1\%$ to $10\%$.  This is completely
incompatible with the observed limits
\cite{Coleman.Glashow,signal.CK,tests}
that $c$ is constant at a fractional level well below $10^{-20}$.

This result is a direct consequence of well-known properties of
quantum field theories, of which the standard model is only one
example.  Indeed the actual technical result is
related to results that can
be found more-or-less explicitly in several of the cited
references --- e.g., by Kosteleck{\'y} and Potting
\cite{Kostelecky:1994rn}. However, what appear to be almost uniformly
missed or 
unemphasized are the consequences that we deduce.  For example,
Coleman and Glashow \cite{Coleman.Glashow} and Colladay and
Kosteleck{\'y} \cite{signal.CK,kostelecky1} recognize that the
dominant effects of Lorentz violation at the Planck scale are
represented by renormalizable \cite{KLP} terms of dimension 4 or less
in the 
standard model considered as a low-energy effective theory.
Therefore they consider a Lorentz-violating extension of the
standard model, and they set some of the observational bounds on
the Lorentz violating parameters. However, 
we frequently see prominent
statements, such as the one in \cite{Kostelecky:1994rn}, that
Lorentz violation is ``likely to be suppressed by some power
of \dots $m_{\rm ew}/M_{\rm Pl} \simeq 10^{-17}$,'' where $m_{\rm
ew}$ is the electroweak scale.  While this assumption is necessary
to agree with experiment, it begs the question of whether the
power suppression actually happens in a given theory with
Planck-scale Lorentz violation. 

Another example is the paper by Burgess et al.\ \cite{Burgess:2002tb},
who consider an effective field theory that arises from a higher
dimensional scenario with Lorentz violation.  At lowest order, Lorentz
violation appears in the effective Lagrangian only in higher dimension
graviton operators. Loop corrections involving gravitons have
Lorentz-violating power-law divergences.  These authors set the
power-law divergences to zero, one justification for which is the use
of dimensional regularization to define the effective theory.  This
appears to give only power-suppressed Lorentz violation, in
contradiction with our estimates.  However, one must also consider
higher order contributions to the coefficients
that relate the effective Lagrangian to the exact theory.  
Corresponding to the power-law
divergences in the effective theory are finite contributions to
coefficients, including those of Lorentz-violating operators of
dimension 4.  The natural size of the Lorentz violation in their
scenario is therefore the same as ours.

Finally, Myers and Pospelov \cite{Myers.Pospelov} state an argument
close to ours, but in the framework of effective field theory, with
tree-approximation Lorentz violation given by higher dimension
operators.  They observe that loop corrections in the effective field
theory give large Lorentz violation in accordance with our estimates.
They then completely reverse their case by claiming to find a
condition in which the Lorentz violation disappears at leading power.
However, as explained by two of us \cite{Perez.Sudarsky}, their
argument does not work; even this second paper failed to take the next
step, which is to realize that since a Planck-scale Lorentz violation
naturally leads to violations of observable Lorentz symmetry at the
percent level, the considered scenarios are ruled out.

A closely related case, to which our arguments apply, is that of
non-commutative field theories (NCFT), which may arise in a suitable
limit from string theory \cite{Seiberg.Witten}.  In NCFT, a modified
structure of space-time (``non-commutativity'') appears as a
Lorentz-violating non-locality in the action defining the field
theory, with a characteristic scale $M$, which might be comparable
with the Planck scale.  Loop corrections in NCFT badly violate Lorentz
invariance \cite{Minwalla:1999px}.  Our general argument applies to
NCFT. Indeed, if NCFT is used without any extra UV cutoff, then loop
momenta extend far above the non-commutativity scale in a highly
Lorentz-violating manner.  Thus these theories suffer from even more
Lorentz violation than we have estimated.  Indeed Anisimov \emph{et
  al.} \cite{ABDG} have shown that NCFT (at least with a
$c$-number non-commutativity tensor $\theta_{\mu\nu}$) are ruled out
unless the UV cutoff $\Lambda$ is nine or ten orders of magnitude
\emph{below} the non-commutativity scale.  
Carlson, Carone and Lebed \cite{CCL}
have set even tighter limits in non-commutative QCD.  Now the raison
d'{\^e}tre of NCFT is to take account of non-standard space-time
structure; but these already-known experimental limits show that NCFT
cease to be applicable at scales far below the non-commutativity
scale.

Notice that our argument has avoided explicit use of the language of
effective field theory.  Conventional treatments of effective field
theory assume a Lorentz invariant cutoff for ultra-violet divergences;
this cutoff is either dimensional regularization or a cutoff on
momenta in \emph{Euclidean} space.  In contrast, we examine the effect
of a effective cutoff in \emph{Minkowski} space, as provided by a
putative granularity of real space-time.

The implications of our argument for both experiment and theory are
quite profound.  First, enthusiasm for improved searches for Lorentz
violation should be dampened; given the stated motivations,
the existing unsuccessful searches suffice by many orders of
magnitude.  Of course, it almost goes without saying that it is
correct science to question and test accepted principles; on these
grounds, tests of Lorentz invariance are worthwhile.  But further
enthusiasm has been engendered by estimates of specific small orders
of magnitude for Lorentz violation.  It is this excess enthusiasm that
we wish to dampen.

As to theory, the critical task concerns any proposal in which Lorentz
symmetry is substantially broken at the Planck scale.  It is to find
and implement a mechanism to give automatic local Lorentz invariance
at low energies, despite a violation at the Planck scale.  We assume
here that the treatment involves real time, not an analytic
continuation to imaginary time, as is common in treatments of quantum
field theories in \emph{flat} space-time.  One mechanism is to have a
custodial symmetry that is sufficient to prohibit Lorentz-violating
dimension 4 terms, without itself being the full Lorentz group.  But
such a symmetry does not appear to be known.  Corresponding issues
arise in any proposal that involves modified dispersion relations,
e.g., \cite{DSR1,DSR2,DSR3}; its proponents must show that the
proposal survives experimental scrutiny after inclusion of known
interactions.

Effectively we have shown that Lorentz invariance should be added to
the well-known list of fine-tuning problems in the standard model; to
date, this list is normally considered to include only the
cosmological constant, the Higgs bare mass, and mass hierarchies.

There is not necessarily a conflict between discreteness and the
absence of a preferred frame.  In \cite{dowker} Dowker et al.\ show
how, by the use of a random causal set of points, space-time can be
made discrete while Lorentz invariance is preserved in a suitable
sense. From a more quantum-theoretic viewpoint, Rovelli and Speziale
\cite{carlo} argue that the existence of a minimum
measurable length does not of itself imply that local Lorentz
invariance is violated any more than the discreteness of the
\emph{eigenvalues} of the angular momentum operators implies
violation of rotational invariance in ordinary quantum mechanics.

An optimistic point of view should be stressed: a branch of
theoretical physics long considered to suffer
from detachment from experimental guidance is now in the
opposite situation.  Because of mechanisms intimately tied to the
known ultra-violet divergences in conventional quantum field theories,
certain kinds of Planck-scale phenomena, like a preferred frame,
manifest themselves suppressed by no powers of energy relative to the
Planck energy, but only by two powers of standard model couplings.
Lorentz invariance continues to play the powerful role it has played
throughout the twentieth century of imposing stringent requirements on
the kinds of mathematical theory that are permitted to agree with
experiment.


\paragraph{Note added:}
After this paper was completed, a paper by
Myers and Pospelov \cite{Myers.Pospelov.2} appeared that contains a
clear statement of the fine-tuning problem for Lorentz invariance.
They deduce that the low-energy theory must have a cutoff far below
the Planck scale, just as in our discussion of NCFT.  However, they do
not explain the problems of implementing such a cutoff
Lorentz-invariantly in Minkowski space, as is appropriate in a theory
of dynamical space-time.

Notice that the cutoff forms an upper limit to the validity of
ordinary quantum field theory.  If physics above the cutoff were
represented by an ordinary quantum field theory, our argument would
apply to that extended theory.


\begin{acknowledgments}
  We would like to thank A. Ashtekar, J.  Banavar, O. Bertolami, C.
  Burgess, Y. Chen, L. Freidel, L. Frankfurt, M. Graesser, A. Guijosa,
  J. Hall, T.  Jacobson, J. Jain, C. Kozameh, A. Kosteleck{\'y}, R.
  Montemayor, M.  Mondrag{\'o}n, H.A.  Morales-T{\'e}cotl, H. Sahlmann, C.
  Stephens, and M.  Strikman for useful discussions and comments. JC
  was supported in part by the U.S. Department of Energy.  LFU
  acknowledges support from projects DGAPA-IN11700 and
  CONACYT-40745-F. DS acknowledges support from the projects
  DGAPA-IN112401 and  IN108103-3 and CONACyT.
  AP acknowledges support from 
  NSF grant PHY-0090091 and the Eberly Research Funds of Penn State
  University. 

\end{acknowledgments}


\appendix*
\section{Estimate of one-loop Lorentz violation}

\setcounter{equation}{0}
\renewcommand\theequation{A.\arabic{equation}}

We now give some details of how Planck-scale Lorentz violation
manifests itself in the propagation of low-energy particles, with a
justification of our estimate of a percent for the observable effects.
As a concrete example, we consider the self-energy graph of a scalar
field with one fermion loop, Fig.\ \ref{fig:loop}, in a Yukawa theory.
The main ideas are quite familiar
and the general considerations involve only
symmetry arguments and normal ultra-violet power-counting, which apply
generally to all self-energy graphs.

Let $\Pi_1(p)$ represent the value of the graph.  We quantify the
dominant Lorentz violation at low energy by the quantity $\xi=
\partial^2\Pi_1(p)/\partial (p^0)^2 + \partial^2\Pi_1(p)/\partial
(p^1)^2$ at $p=0$; this would be 
zero for a Lorentz-invariant self-energy.

Without the Planck-scale modifications, we would have
\begin{equation}
   \xi
   = -\frac{ ig^2 }{ \pi^4 }
   \int \diff{^4k}
   \,\,
   \frac{ [(k^0)^2 + (k^1)^2]\, (k^2+3M^2)  }
      { (k^2-M^2+i\epsilon)^4 }.
\end{equation}
Were it not that this integral is logarithmically divergent, it could
be shown to be zero by continuing $k^0$ to imaginary values and then
using Euclidean rotation invariance.

When there is Lorentz-violation in the free fermion propagator, as in
\cite{calcs,lqg1},
the
integrand would only apply in the low-energy limit.  When $k$ is at or
above the Planck scale, there would be substantial and
Lorentz-violating modifications.

The would-be logarithmic divergence ensures that there is an order
unity contribution from Planck-scale momenta, with their Lorentz
violation.  Note that a cutoff provided by preferred frame granularity
is Lorentz-violating.  As a simple illustration we modify the usual
free-fermion propagator $i(\gamma\cdot k + m)/(k^2-m^2+i\epsilon)$ by
a factor of a 
smooth function $f(|\ve{k}|/\Lambda)$ that obeys $f(0)=1$ and
$f(\infty)=0$, with 
a cutoff parameter $\Lambda$.  Then
\begin{equation}
   \xi = \frac{g^2} {6\pi^2}
 \left[ 1 + 2 \int \limits_0^{\infty} \diff{x} x f'(x)^2 \right] .
\end{equation}
Thus the corresponding Lorentz violation is of order the square of the
coupling independently of $\Lambda$.  The exact value depends
on the details of the Planck-scale free propagator, of course.  The
main point, however, is that the power counting that gives the
would-be logarithmic divergence follows from standard arguments in the
theory of renormalization, and that it applies to self-energy graphs
for all fields.  So the above integral gives a reasonable estimate of
the size of the Lorentz violation, in the absence of some special
cancellation. Typical measured standard-model couplings then give our
percent estimate.

By use of (\ref{eq:inv.prop}), it can be checked that the $\xi$
coefficient corresponds to a modification of the dispersion
relation (\ref{eq:DR.normal}) in which the speed of light
parameter $c$ is modified by a factor $1+\xi/4 + O(\xi^2)$. In a
theory with one field it is possible to treat this term as a
renormalization of the space-time metric tensor which could remove
the observable Lorentz violation. However, there are many fields
in the standard model that differ by the sizes of their couplings.
Hence renormalization of the metric tensor cannot 
remove all leading-power Lorentz violation.

Given a specific form of space-time granularity, accurate quantitative
predictions could be obtained, with the use of renormalization-group
methods etc.  But such details would not affect the drastic
disagreement of the percent estimate with actual observations.


\end{document}